\documentclass[a4paper, 12pt, oneside]{article}
\usepackage[cp1251]{inputenc}
\usepackage[english, russian]{babel}
\usepackage{ graphics,amsfonts, amsmath,bm,amssymb, array}
\usepackage[dvips]{graphicx}
\usepackage[flushleft,small]{caption2}
\renewcommand{\Re}{\mathop{\rm Re\,}}
\renewcommand{\Im}{\mathop{\rm Im\,}}
\renewcommand{\max}{\mathop{\rm max\,}}

\makeatother {\renewcommand{\baselinestretch}{1.15}

\begin{document}
\thispagestyle{empty} \large
\renewcommand{\abstractname}{\, }
\renewcommand{\refname}{\begin{center} REFERENCES\end{center}}

\begin{flushright}\it\large
Dedicated to the Memory\\ of our Teachers\\ C. Cercignani and
K. Case
\end{flushright}

 \begin{center}
\bf Аналитическое решение второй задачи Стокса о поведении газа
над колеблющейся поверхностью. Часть III: решение задачи и приложения
\end{center}\medskip
\begin{center}
  \bf V. A. Akimova\footnote{$vikont\_ava@mail.ru$},
  A. V. Latyshev\footnote{$avlatyshev@mail.ru$} and
  A. A. Yushkanov\footnote{$yushkanov@inbox.ru$}
\end{center}\medskip

\begin{center}
{\it Faculty of Physics and Mathematics,\\ Moscow State Regional
University, 105005,\\ Moscow, Radio str., 10--A}
\end{center}\medskip

\tableofcontents
\setcounter{secnumdepth}{4}

\begin{abstract}
В настоящей работе сформулирована и решена аналитически вторая задача Стокса о поведении разреженного газа, заполняющего полупространство. Плоскость,
ограничивающая полупространство, совершает гармонические колебания
в своей плоскости. Используется кинетическое уравнение с модельным интегралом столкновений в форме $\tau$--модели. Рассматривается случай диффузного отражения молекул газа от стенки. Построена функция распределения газовых молекул, найдена массовая скорость газа в полупространстве, отыскивается ее значение непосредственно у стенки. Найдена сила сопротивления, действующая со стороны газа на границу, совершающую в своей плоскости колебательное движение. Кроме того, отыскивается мощность диссипации энергии, приходящаяся на единицу площади колеблющейся пластины, ограничивающей газ.


\end{abstract}

\begin{center}
\item{}\section{Введение}
\end{center}

Задача о поведении газа над движущейся поверхностью в последние годы
привлекает пристальное внимание \cite{Stokes} -- \cite{15}. Это связано с развитием современных технологий, в частности, технологий наноразмеров.
В \cite{Yakhot} -- \cite{15} эта задача решалась численными или
приближенными методами. В настоящей работе показано, что эта задача допускает аналитическое решение. Аналитическое решение строится с помощью теории обобщенных функций и сингулярных интегральных уравнений.

В настоящей работе построено аналитическое решение второй задачи Стокса. На основе аналитического решения вычисляется скорость газа в полупространстве и непосредственно у колеблющейся границы, найдена сила трения, действующая со стороны газа на колеблющуюся пластину, а также находится диссипация энергии пластины.
\begin{center}
\item{}\subsection{История проблемы}
\end{center}

Впервые задача о поведении газа над стенкой, колеблющейся в своей плоскости, была рассмотрена Дж. Г. Стоксом \cite{Stokes}. Задача решалась гидродинамическим методом без учёта эффекта скольжения. Обычно такую задачу называют второй задачей Стокса \cite{Yakhot}--\cite{SS-2002}.

В последние годы на тему этой задачи появился ряд публикаций. В работе \cite{Yakhot} задача рассматривается для любых частот колебания поверхности. Из кинетического уравнения БГК получено уравнение типа гидродинамического. Рассматриваются гидродинамические граничные условия. Вводится коэффициент, связывающий скорость газа на поверхности со скоростью поверхности. Показано, что в случае высокочастотных колебаний сила трения, действующая на поверхность, не зависит от частоты.

В работе \cite{SK-2007} получены коэффициенты вязкостного и теплового скольжения с использованием различных модельных уравнений. Использованы как максвелловские граничные условия, так и граничные условия Черчиньяни --- Лэмпис.

В статье \cite{10} рассматривается газовый поток над бесконечной пластиной, совершающей гармонические колебания в собственной плоскости. Найдена скорость газа над поверхностью и сила, действующая на поверхность со стороны газа. Для случая низких частот задача решена на основе уравнения Навье --- Стокса. Для произвольных скоростей колебаний поверхности задача решена численными методами на основе кинетического уравнения Больцмана с интегралом столкновений в форме БГК (Бхатнагар, Гросс, Крук). При этом рассматривался только случай чисто диффузного отражения молекул от поверхности. Дано аналитическое решение для случая колебаний высокой частоты.

Работа \cite{11} является экспериментальным исследованием. Изучается поток газа, создаваемый механическим резонатором при различных частотах колебания резонатора. Эксперименты показывают, что при низких частотах колебаний резонатора, действующая на него со стороны газа сила трения прямо пропорциональна частоте колебания резонатора. При высоких частотах колебания резонатора ($~10^8$ Гц) действующая на него сила трения от частоты колебаний не зависит.

В последнее время задача о колебаниях плоской поверхности в собственной плоскости изучается и для случая неньютоновских жидкостей \cite{5} и
\cite{6}.

В статье \cite{12} рассматривается пример практического применения колебательной системы, подобной рассматриваемой во второй задаче Стокса, в области нанотехнологий.

Общим существенным недостатком всех упомянутых теоретических работ по решению второй задачи Стокса является отсутствие учёта характера взаимодействия с поверхностью, т.е. рассматривается только случай полной аккомодации тангенциального импульса.

Коэффициент аккомодации тангенциального импульса является величиной, зависящей от состояния поверхности. И если в "естественном"\, состоянии значение этой величины как правило близко к единице, то при специальной обработке поверхности её значение можно уменьшить многократно \cite{13}, а значит и существенно изменить характер взаимодействия поверхности с прилегающим газом.

В диссертации \cite{15} были предложены два решения второй задачи Стокса, учитывающие весь возможный диапазон коэффициента аккомодации тангенциального импульса. Эти решения отвечают соответственно гидродинамическому и кинетическому описанию поведения газа над колеблющейся поверхностью в режиме со скольжением.
В конце второй главы диссертации \cite{15} проведено сопоставление с результатами, полученными в статье \cite{10}.

В наших работах \cite{ALY-1} и \cite{ALY-2} для второй задачи Стокса
отыскиваются собственные функции и соответствующие собственные значения, отвечающие как дискретному, так и непрерывному спектрам. Исследована структура дискретного и непрерывного спектров. Решена краевая задача Римана из теории функций комплексного переменного, лежащая в основе аналитического решения второй задачи Стокса. Развивается математический аппарат, необходимый для аналитического решения задачи и приложений.

В настоящей работе строится аналитическое решение второй задачи Стокса. На основе аналитического решения вычисляется скорость газа в полупространстве и непосредственно у колеблющейся границы, найдена сила трения, действующая со стороны газа на колеблющуюся пластину, а также находится диссипация энергии пластины.

\begin{center}
\item{}\subsection{Содержание работы}
\end{center}

В п. 2 рассматривается постановка второй задачи Стокса. Задача формулируется в общей постановке --- с использованием граничных условий Максвелла (зеркально -- диффузных граничных условий). Далее задача будет рассматриваться только для диффузных граничных условий.

В качестве кинетического уравнения рассматривается линеаризованное кинетическое уравнение. Это уравнение получается путем линеаризации модельного кинетического уравнения Больцмана и интегралом столкновений в форме релаксационной $\tau$--модели.

Пластина (плоскость), ограничивающая полупространство с разреженным газом совершает колебательные движения вдоль оси $y$. В качестве граничных условий используются два условия. Одно из них --- граничное условие вдали от стенки --- требует исчезания функции $h(x_1,\mu)$ вдали от стенки. Второе условие --- условие на стенке --- вытекает из требования диффузного отражения молекул от стенки.

Требуется определить функцию распределения газовых молекул, найти скорость газа в полупространстве и непосредственно у стенки, найти силу трения, действующую со стороны газа на пластину, найти мощность диссипации энергии пластины.

В п. 3 кинетическое уравнение упрощается путем представления функции распределения в виде произведения $y$--компоненты скорости молекул газа на новую неизвестную функцию. При этом получается однопараметрическое семейство кинетических уравнений с чисто мнимым параметром. Параметром уравнений служит безразмерная величина частоты колебаний пластины. Эта величина $\omega_1=\omega/\nu=\omega \tau$ равна частоте колебаний пластины $\omega$, деленной на величину частоты $\eta$ столкновений молекул газа, $\tau=1/\nu$ -- время между двумя последовательными столкновениями молекулы.

В п. 3 приводятся собственные решения (непрерывные моды) исходного кинетического уравнения, отвечающие непрерывному и дискретному спектрам.
Вводится коэффициент задачи. Под коэффициентом $G(\mu)$ задачи понимается отношение граничных значений
дисперсионной функции сверху и снизу на действительной оси:
$G(\mu)=\lambda^+(\mu)/\lambda^-(\mu)$. Выясняется, что существует критическая частота
$$
\omega_1^*=\max\limits_{0<\mu<+\infty}\sqrt{[\Im\lambda^+(\mu)]^2-
[\Re\lambda^+(\mu)]^2}\approx 0.733,
$$
такая, что при $\omega_1\in[0,\omega_1^*)$ индекс коэффициента задачи равен единице: $\varkappa(G)=1$, а при $\omega_1\in (\omega_1^*,+\infty)$ индекс коэффициента задачи равен нулю: $\varkappa(G)=0$. Таким образом, если $\omega_1$ находится в первом (левом) регионе, то дисперсионная функция имеет два комплексно -- значных нуля, отличающихся лишь знаками в силу четности дисперсионной функции. Если параметр $\omega_1$ находится во втором (правом) регионе, то индекс задачи равен нулю, т.е. дисперсионная функция комплексно -- значных нулей не имеет.

В п. 4 и п. 5 строится аналитическое решение поставленной граничной задачи. Решение ищется в виде суммы собственной дискретной моды, умноженной на неизвестную постоянную (коэффициент дискретного спектра) и интеграла от собственных непрерывных мод, умноженных на неизвестную функцию (коэффициент непрерывного спектра). Это разложение решения, автоматически удовлетворяющее граничному условию вдали от стенки, подставляется в граничное условие на стенке.

Получается сингулярное интегральное уравнение с ядром Коши. Путем введения неизвестной функции (типа интеграла Коши) сингулярное уравнение сводится к неоднородной краевой задаче Римана, соответствующая однородная задача которой рассмотрена выше.

Решение неоднородной краевой задачи Римана ищется в классе исчезающих в бесконечно удаленной точке функций в случае $\varkappa(G)=1$, и в классе ограниченных в беконечно удаленной точке функций в случае $\varkappa(G)=0$.

С помощью решения задачи Римана находятся коэффициенты разложения решения исходной краевой задачи, отвечающие дискретному и непрерывному спектрам.

Аналитическое решение построено.

В п. 6 находится скорость разреженного газа в полупространстве и непосредственно у стенки. Существенно используется интегральное представление функции факторизующей функции и обратной к ней величины.

Затем в п. 7 исследуется гидродинамический характер решения. Показано, что при малых скоростях ограничивающей газ плоскости решение задачи переходит в известное решение из механики сплошной среды.

В п. 8 находятся сила трения, действующая со стороны газа на пластину, и мощность диссипации энергии пластины.

\begin{center}
\item{}\section{Линеаризованное кинетическое уравнение для задачи о колебаниях газа}
\end{center}

Пусть разреженный одноатомный газ занимает полупространство $x>0$
над плоской твердой поверхностью, лежащей в плоскости $x=0$.
Поверхность $(y,z)$ совершает гармонические колебания вдоль оси $y$
по закону $u_s(t)=u_0\cos \omega t$.

Рассмотрим линеаризованное кинетическое уравнение
$$
\dfrac{\partial \varphi}{\partial t}+v_x\dfrac{\partial \varphi}{\partial x}+\varphi(x,t,\mathbf{v})=\dfrac{\nu m}{kT}v_yu_y(x,t).
\eqno{(1.1)}
$$

В (1.1) $\nu=1/\tau$ -- частота столкновений газовых молекул, $\tau$ -- время между двумя последовательными столкновениями молекул, $m$ -- масса молекулы, $k$ -- постоянная Больцмана, $T$ --
температура газа, $u_y(x)$ -- массовая скорость газа,
$$
u_y(x,t)=\dfrac{1}{n}\int f(x,t,\mathbf{v})d^3v,
\eqno{(1.2)}
$$
$n$ -- числовая плотность (концентрация) газа. Концентрация газа и его температура считаются постоянными в линеаризованной постановке задачи.

Введем безразмерные скорости и параметры: безразмерную скорость молекул:
$\mathbf{C}=\sqrt{\beta}\mathbf{v}$ \;$(\beta=m/(2kT))$, безразмерную массовую скорость $U_y(x,t)=\sqrt{\beta}u_y(x,t)$, безразмерное время $t_1=\nu t$ и
безразмерную скорость колебаний пластины $U_s(t)=U_0\cos\omega t$,
где $U_0=\sqrt{\beta}u_0$ -- безразмерная амплитуда скорости колебаний границы полупространства. Тогда уравнение (1.1) может быть записано в виде:
$$
\dfrac{\partial \varphi}{\partial t_1}+C_x\dfrac{\partial \varphi}{\partial x_1}+\varphi(x_1,t_1,\mathbf{C})={2C_y}U_y(x_1,t_1).
\eqno{(1.3)}
$$

Заметим, что для безразмерного времени $U_s(t_1)=U_0\cos\omega_1t_1$.

В задаче о колебаниях газа требуется найти функцию распределения $f(x_1,t_1,\mathbf{C})$ газовых молекул. Функция распределения свзана
с функцией $\varphi(x_1,t_1,C_x)$ соотношением:
$$
f(x_1,t_1,\mathbf{C})=f_M(C)\big[1+\varphi(x_1,t_1,C_x)\big],
\eqno{(1.4)}
$$
где
$$
f_M(C)=n\Big(\dfrac{\beta}{\pi}\Big)^{3/2}\exp(-C^2)
$$
-- есть абсолютный максвеллиан.

Затем на основании найденной функции распределения требуется найти массовую скорость газа, значение
массовой скорости газа непосредственно у стенки. Кроме того, требуется вычислить силу сопротивления газа, действующую на колеблющуюся пластину, ограничивающую газ.
Подчеркнем, что задача о колебаниях газа решается в линеаризованной постановке.
Линеаризация задачи проведена по безразмерной массовой скорости $U_y(x_1,t_1)$ при условии, что $|U_y(x,t_1)|\ll 1$. Это неравенство эквивалентно неравенству
$$
|u_y(x_1,t_1)|\ll v_T,
$$
где $v_T=1/\sqrt{\beta}$ -- тепловая скорость молекул, имеющая порядок скорости звука.

Величину безразмерной массовой скорости $U_y(t_1,x_1)$ найдем из ее определения (1.2)
$$
U_y(x_1,t_1)=\dfrac{1}{\pi^{3/2}}\int \exp(-C^2)C_y\varphi(x_1,t_1,
\mathbf{C})d^3C.
\eqno{(1.5)}
$$

С помощью (1.5) кинетическое линеаризованное уравнение (1.3) записывается в виде:
$$
\dfrac{\partial \varphi}{\partial t_1}+C_x\dfrac{\partial \varphi}{\partial x_1}+\varphi(x_1,t_1,\mathbf{C}) =\dfrac{2C_y}{\pi^{3/2}}
\int\exp(-{C'}^2)C_y'\varphi(x_1,t_1,\mathbf{C'})\,d^3C'.
\eqno{(1.6)}
$$

Сформулируем зеркально--диффузные граничные условия, записанные относительно функции $\varphi(x_1,t_1,\mathbf{C})$:
$$
\varphi(0,t_1,\mathbf{C})=2qC_yU_s(t_1)+(1-q)\varphi(0,t_1,
-C_x,C_y,C_z),\quad C_x>0,
\eqno{(1.7)}
$$
и
$$
\varphi(x_1\to+\infty,t_1,\mathbf{C})=0.
\eqno{(1.8)}
$$

Итак, граничная задача о колебаниях газа сформулирована полностью и состоит в решении уравнения (1.6) с граничными условиями (1.7) и (1.8).

Отметим, что к выражению (1.5) для безразмерной массовой скорости можно придти, исходя из определения размерной массовой скорости газа (1.2). В самом деле, подствляя в (1.2) выражение (1.4), приходим в точности к выражению (1.5).

\begin{center}
\item{}\section{Декомпозиция граничной задачи}
\end{center}

Учитывая, что колебания пластины рассматриваются вдоль оси $y$, будем искать, следуя Черчиньяни \cite{16}, функцию $\varphi(x_1,t_1,\mathbf{C})$ в виде
$$
\varphi(x_1,t_1,\mathbf{C})=C_yH(x_1,t_1,C_x).
\eqno{(2.1)}
$$
Тогда безразмерная массовая скорость (1.5) с помощью (2.1) равна
$$
U_y(x_1,t_1)=\dfrac{1}{2\sqrt{\pi}}\int\limits_{-\infty}^{\infty}
\exp(-C_x'^2)H(x_1,t_1, C_x')dC_x'.
\eqno{(2.2)}
$$

С помощью указанной выше подстановки (2.1) кинетическое уравнение (1.6) преобразуется к виду:
$$
\dfrac{\partial H}{\partial t_1}+C_x\dfrac{\partial H}{\partial x_1}+
H(x_1,t_1,C_x)=\dfrac{1}{\sqrt{\pi}}\int\limits_{-\infty}^{\infty}
\exp(-C_x'^2)H(x_1,t_1,C_x')dC_x'.
\eqno{(2.3)}
$$

Граничные условия (1.7) и (1.8) преобразуются в следующие:
$$
H(0,t_1,C_x)=2qU_s(t_1)+(1-q)H(0,t_1,-C_x),\qquad C_x>0,
\eqno{(2.4)}
$$
$$
H(x_1\to +\infty,t_1, C_x)=0.
\eqno{(2.5)}
$$

Следующим шагом одновременно осуществим комплексификацию кинетического
уравнения и выделим временную переменную, положив далее:
$$
H(x_1,t_1,C_x)=\Re\{e^{-i\omega_1t_1}h(x_1,C_x)\}
\eqno{(2.6)}
$$
и
$$
U_0\cos\omega_1t_1=\Re\{e^{-i\omega_1t_1}U_0\}.
$$

Теперь мы получаем комплексно--значное уравнение (уравнение относительно
комплексно--значной функции $h(x_1,C_x)$):
$$
C_x\dfrac{\partial h}{\partial x_1}+(1-i\omega_1)h(x_1,C_x)
=\dfrac{1}{\sqrt{\pi}}\int\limits_{-\infty}^{\infty}
\exp(-C_x'^2)h(x_1,C_x')dC_x'.
\eqno{(2.7)}
$$

Граничные условия (2.4) и (2.5) переходят в следующие:
$$
h(0,C_x)=2qU_0+(1-q)h(0,-C_x),\qquad C_x>0,
\eqno{(2.8)}
$$
и
$$
h(x_1\to+\infty,C_x)=0.
\eqno{(2.9)}
$$

Тогда безразмерная массовая скорость равна:
$$
U_y(t_1,x_1)=\dfrac{1}{2\sqrt{\pi}}\int\limits_{-\infty}^{\infty}
\exp(-C_x'^2)\Re\{e^{-i\omega_1t_1}h(x_1,C_x')\}dC_x'.
\eqno{(2.10)}
$$

Мы получили граничную задачу, состоящую в решении уравенния (2.7)
с граничными условиями (2.8) и (2.9).
Далее будем рассмтривать задачу с диффузными граничными условиями.
Перепишем граничную задачу (2.7), (2.8) и (2.9) в виде:
$$
\mu\dfrac{\partial h}{\partial x_1}+z_0h(x_1,\mu)=\dfrac{1}{\sqrt{\pi}}
\int\limits_{-\infty}^{\infty}\exp(-{\mu'}^2)h(x_1,\mu')d\mu',
\eqno{(2.11)}
$$
где
$$
z_0=1-i\omega_1,
$$
и
$$
h(0,\mu)=2U_0,\qquad \mu>0,
\eqno{(2.12)}
$$
$$
h(+\infty,\mu)=0.
\eqno{(2.13)}
$$

Разделение переменных в уравнении (2.11) осуществляется следующей подстановкой
$$
h_\eta(x_1,\mu)=\exp\Big(-\dfrac{x_1z_0}{\eta}\Big)\Phi(\eta,\mu),
\eqno{(2.14)}
$$
где $\eta$ -- параметр разделения, или спектральный параметр, вообще говоря, комплексный.

Подставляя (2.14) в уравнение (2.11) получаем характеристическое уравнение
$$
(\eta-\mu)\Phi(\eta,\mu)=\dfrac{\eta}{\sqrt{\pi}z_0}
\int\limits_{-\infty}^{\infty}
\exp(-{\mu'}^2)\Phi(\eta,\mu')d\mu'.
\eqno{(2.15)}
$$
Если ввести обозначение
$$
n(\eta)=\dfrac{1}{z_0}\int\limits_{-\infty}^{\infty}
\exp(-{\mu'}^2)\Phi(\eta,\mu')d\mu',
\eqno{(2.16)}
$$
то уравнение (2.15) может быть записано с помощью (3.6) в виде
$$
(\eta-\mu)\Phi(\eta,\mu)=\dfrac{1}{\sqrt{\pi}}\eta n(\eta),\qquad
\eta\in \mathbb{C}.
\eqno{(2.17)}
$$

Решение характеристического уравнения для действительных значений
параметра $\eta$ будем искать в пространстве
обобщенных функций \cite{6}.
Обобщенное решение уравнения (2.17) при $n(\eta)\equiv 1$ имеет вид:
$$
\Phi(\eta,\mu)=\dfrac{1}{\sqrt{\pi}}\eta n(\eta)P\dfrac{1}{\eta-\mu}+
e^{\eta^2}\lambda(\eta)\delta(\eta-\mu),
\eqno{(2.18)}
$$
где $-\infty<\eta, \mu <+\infty$.

Здесь $\delta(x)$ -- дельта--функция Дирака, символ $Px^{-1}$
означает главное значение интеграла при интегрировании $x^{-1}$,
$\lambda(z)$ -- дисперсионная функция, введенная равенством
$$
\lambda(z)=1-i\omega_1+\dfrac{z}{\sqrt{\pi}}\int\limits_{-\infty}^{\infty}
\dfrac{\exp(-\tau^2)d\tau}{\tau-z}.
$$
Эту функцию можно преобразовать к виду: $\lambda(z)=-i\omega_1+\lambda_0(z)$,
где $\lambda_0(z)$ -- известная функция из теории плазмы,
$$
\lambda_0(z)=\dfrac{1}{\sqrt{\pi}}\int\limits_{-\infty}^{\infty}
\dfrac{e^{-\tau^2}\tau d\tau}{\tau-z}.
$$

Собственные функции (2.18) называются собственными функциями непрерывного
спектра, ибо спектральный параметр $\eta$ непрерывным образом заполняет всю действительную прямую.

Таким образом, собственные решения уравнения (2.11) имеют вид
$$
h_\eta(x,\mu)=\exp\Big(-\dfrac{x_1}{\eta}z_0\Big)
\Big[\dfrac{1}{\sqrt{\pi}}\eta P\dfrac{1}{\eta-\mu}+
\exp(\eta^2)\lambda(\eta)\delta(\eta-\mu)\Big].
\eqno{(2.19)}
$$

Собственные решения (2.19) отвечают непрерывному спектру характеристического
уравнения, ибо спектральный параметр непрерывным образом пробегает всю числовую прямую, т.е. непрерывный спектр $\sigma_c$
есть вся конечная часть числовой прямой: $\sigma_c=(-\infty,+\infty)$.

По условию задачи мы ищем решение, невозрастающее вдали от стенки.
Поэтому далее будем рассматривать положительную часть непрерывного спектра. В этом случае собственные решения (2.19) являются исчезающими вдали от стенки. В связи с этим спектром граничной задачи будем называть положительную действительную полуось параметра $\eta$:
$\sigma_c^{\rm problem}=(0,+\infty)$.

Приведем формулы Сохоцкого для дисперсионной функции:
$$
\lambda^{\pm}(\mu)=\pm i\sqrt{\pi}\mu e^{-\mu^2}-i\omega_1+
\dfrac{1}{\sqrt{\pi}}\int\limits_{0}^{\infty}
\dfrac{e^{-\tau^2}\tau d\tau}{\tau-\mu}.
$$
Разность граничных значений дисперсионной функции отсюда равна:
$$
\lambda^+(\mu)-\lambda^-(\mu)=2\sqrt{\pi}\mu e^{-\mu^2}i,
$$
полусумма граничных значений равна:
$$
\dfrac{\lambda^+(\mu)+\lambda^-(\mu)}{2}=-i\omega_1+\dfrac{1}{\sqrt{\pi}}
\int\limits_{0}^{\infty}\dfrac{e^{-\tau^2}\tau d\tau}{\tau-\mu}.
$$

Заметим, что на действительной оси действительная часть
дисперсионной функции $\lambda_0(\mu)$ имеет два нуля $\pm\mu_0$, $\mu_0=0.924\cdots$. Эти два нуля в силу четности функции $\lambda_0(\mu)$ различаются лишь знаками.

Отметим, что на действительной оси дисперсионную функцию удобнее использовать в численных расчетах в виде (см. \cite{19})
$$
\lambda_0(\mu)=1-2\mu^2 \int\limits_{0}^{1}\exp(-\mu^2(1-t^2))dt,\qquad
\mu\in(-\infty,+\infty).
$$

Разложим дисперсионную функцию в ряд Лорана по отрицательным степеням
переменного $z$ в окрестности бесконечно удаленной точки:
$$
\lambda(z)=-i\omega_1-\dfrac{1}{2z^2}-\dfrac{3}{4z^4}-\dfrac{15}{8z^6}-\cdots,
\quad z\to \infty.
\eqno{(2.20)}
$$

Из разложения (2.20) видно, что при малых значениях $\omega_1$
дисперсионная функция имеет два отличающиеся лишь знаками комплексно--значных нуля:
$$
\pm\eta_0^{(0)}(\omega_1)=\dfrac{1+i}{2\sqrt{\omega_1}}.
$$

Отсюда видно, что при $\omega_1\to 0$ оба нуля дисперсионной функции
имеют пределом одну бесконечно удаленную точку $\eta_i=\infty$ кратности (порядка) два.

Из разложения (2.20) видно так же, что значение дисперсионной функции в
бесконечно удаленной точки равно:
$$
\lambda(\infty)=-i\omega_1.
$$

Введем выделенную частоту колебаний пластины, ограничивающей газ:
$$
\omega_1^*=\max\limits_{0<\mu<+\infty}\sqrt{-\lambda_0^2(\mu)+s^2(\mu)}\approx 0.733.
$$

Эту частоту колебаний будем называть {\it критической}.

В \cite{ALY-2} показано, что в случае, когда частота колебаний пластины меньше критической, т.е. при $0\leqslant \omega <\omega_1^*$, индекс функции $G(t)$ равен единице. Это означает, что число комплексно--значных нулей дисперсионной функции в разрезанной комплексной плоскости с разрезом вдоль действительной оси, равно двум.

В случае, когда частота колебаний пластины превышает критическую ($\omega>\omega_1^*$) индекс функции $G(t)$ равен нулю: $\varkappa(G)=0$. Это означает, что дисперсионная функция не имеет нулей в верхней и нижней полуплоскостях. В этом случае дискретных (частных) решений исходное кинетическое уравнение (3.1) не имеет.

Таким образом, дискретный спектр характеристического уравнения, состоящий из нулей дисперсионной функции, в случае $0\leqslant \omega_1<\omega_1^*$ есть множество из двух точек $\sigma_d(\omega_1)=\{\eta_0(\omega_1),
-\eta_0(\omega_1)\}$. При $\omega_1>\omega_1^*$ дискретный спектр --- это пустое множество. При $0\leqslant \omega_1<\omega_1^*$ собственными функциями характеристического уравнения являются следующие два решения характеристического уравнения:
$$
\Phi(\pm \eta_0(\omega_1),\mu)=\dfrac{1}{\sqrt{\pi}}\dfrac{\pm \eta_0(\omega_1)}{\pm \eta_0(\omega_1)-\mu}
$$
и два соответствующих собственных решения исходного характеристического уравнения (2.11):
$$
h_{\pm \eta_0(\omega_1)}(x_1,\mu)=\exp \Big(-\dfrac{x_1z_0}{\pm \eta_0(\omega_1)}\Big)\dfrac{1}{\sqrt{\pi}}\dfrac{\pm \eta_0(\omega_1)}{\pm \eta_0(\omega_1)-\mu}.
$$

Под $\eta_0(\omega_1)$ будем понимать тот из нулей дисперсионной функции, который обладает свойством:
$$
\Re \dfrac{1-i\omega_1}{\eta_0(\omega_1)}>0.
$$
Для этого нуля убывающее собственное решение кинетического уравнения (3.1) имеет вид
$$
h_{\eta_0(\omega_1)}(x_1,\mu)=\dfrac{1}{\sqrt{\pi}}
\exp\Big(-\dfrac{x_1z_0}{\eta_0(\omega_1)}\Big)\dfrac{\eta_0(\omega_1)}
{\eta_0(\omega_1)-\mu}.
$$
Это означает, что дискретный спектр рассматриваемой граничной задачи состоит из одной точки $\sigma_d^{\rm problem}=\{\eta_0(\omega_1)\}$ в случае $0 <\omega_1<\omega_1^*$. При $\omega_1\to 0$ оба нуля, как уже указывалось выше, перемещаются в одну и ту же бесконечно удаленную точку. Это значит, что в этом случае дискретный спектр характеристического уравнения состоит из одной бесконечно удаленной точки кратности два:
$\sigma_d(0)=\eta_i=\infty$ и является присоединенным к непрерывному спектру. Этот спектр является также и спектром рассматриваемой граничной задачи. Однако, в этом случае дискретных (частных) решения ровно два:
$$
h_1(x_1,\mu)=1, \qquad h_2(x_1,\mu)=x_1-\mu.
$$

\begin{center}
\item{}\section{Аналитическое решение граничной задачи. Индекс задачи равен нулю}
\end{center}

Составим общее решение уравнения (2.11) в виде интеграла по
непрерывному спектру от собственных решений:
$$
h(x_1,\mu)=\int\limits_{0}^{\infty}
\exp\Big(-\dfrac{x_1}{\eta}z_0\Big)
\Big[\dfrac{1}{\sqrt{\pi}}\eta P\dfrac{1}{\eta-\mu}+
\exp(\eta^2)\lambda(\eta)\delta(\eta-\mu)\Big]a(\eta)d\eta,
\eqno{(3.1)}
$$
или, кратко,
$$
h(x_1,\mu)=\int\limits_{0}^{\infty}
\exp\Big(-\dfrac{x_1}{\eta}z_0\Big)\Phi(\eta,\mu)a(\eta)d\eta.
\eqno{(3.2)}
$$
Здесь $\Phi(\eta,\mu)$ -- собственные функции характеристического уравнения,
отвечающие непрерывному спектру и единичной нормировке,
$$
\Phi(\eta,\mu)=\dfrac{1}{\sqrt{\pi}}\eta P\dfrac{1}{\eta-\mu}+
\exp(\eta^2)\lambda(\eta)\delta(\eta-\mu)
$$
$a(\eta)$ -- неизвестная функция, отвечающая непрерывному спектру.
Эта функция подлежит нахождению из граничных условий (2.12) и (2.13).

Решение (3.1) можно представить в классическом виде:
$$
h(x_1,\mu)=\dfrac{1}{\sqrt{\pi}}\int\limits_{0}^{\infty}
\exp\Big(-\dfrac{x_1z_0}{\eta}\Big)\dfrac{\eta a(\eta)d\eta}{\eta-\mu}+
\exp\Big(-\dfrac{x_1z_0}{\mu}+\mu^2\Big)\lambda(\mu)a(\mu)\theta_+(\mu),
\eqno{(3.3)}
$$
где $\theta_+(\mu)$ -- функция Хэвисайда,
$$
\theta_+(\mu)=\left\{\begin{array}{c}
                       1,\qquad \mu>0, \\
                       0,\qquad \mu<0
                     \end{array}.\right.
$$

Очевидно, что разложение (3.3)
автоматически удовлетворяет граничному условию (2.13) вдали
от стенки. Подставим разложение (3.3) в граничное условие (2.12).
Получаем одностороннее сингулярное интегральное уравнение с ядром Коши
$$
\dfrac{1}{\sqrt{\pi}}\int\limits_{0}^{\infty}
\dfrac{\eta a(\eta)d\eta}{\eta-\mu}+
\exp(\mu^2)\lambda(\mu)a(\mu)\theta_+(\mu)=2U_0,\qquad \mu>0.
\eqno{(3.4)}
$$

Введем вспомогательную функцию
$$
N(z)=\dfrac{1}{\sqrt{\pi}}\int\limits_{0}^{\infty}\dfrac{\eta a(\eta)d\eta}
{\eta-\mu}.
\eqno{(3.5)}
$$

Для этой функции выполняются формулы Сохоцкого:
$$
N^+(\mu)-N^-(\mu)=2\sqrt{\pi}i \mu a(\mu), \qquad \mu>0,
$$
$$
\dfrac{N^+(\mu)+N^-(\mu)}{2}=N(\mu),
$$
где
$$
N(\mu)=\dfrac{1}{\sqrt{\pi}}\int\limits_{0}^{\infty}
\dfrac{\eta a(\eta)d\eta}{\eta-\mu},\qquad \mu>0.
$$

Пользуясь формулами Сохоцкого для вспомогательной и дисперсионной функций, приходим к краевому условию:
$$
\lambda^+(\mu)[N^+(\mu)-2U_0]- \lambda^-(\mu)[N^-(\mu)-2U_0]=0,\qquad
\mu>0.
\eqno{(3.6)}
$$

Уравнение (3.6) --- это краевое условие неоднородной краевой задачи
Римана --- Гильберта. Эта задача состоит в отыскании такой
неизвестной функции $N(z)$, аналитической вдоль разрезанной
плоскости положительной полуоси, граничные значения которой
на берегах этого разреза удовлетворяют краевому условию (3.6).
Рассмотрим соответствующую однородную краевую задачу Римана:
$$
X^+(\mu)=G(\mu)X^-(\mu),\qquad \mu>0,
\eqno{(3.7)}
$$
где
$$
G(\mu)=\dfrac{\lambda^+(\mu)}{\lambda^-(\mu)}.
$$

Решение задачи (3.7) было рассмотрено в \cite{ALY-2}.
Отсутствие нулей дисперсионной функции означает,
что приращение $\ln G(\mu)$ на полуоси $[0,+\infty)$ равно нулю.
Поэтому решение задачи о скачке (3.7) (при $k = 0$) дается интегралом
типа Коши:
$$
X(z)=\exp V(z),
\eqno{(3.8)}
$$
где $V(z)$ понимается как интеграл типа Коши
$$
V(z)\equiv \ln X(z)=\dfrac{1}{2\pi i}\int\limits_{0}^{\infty}
\dfrac{\ln G(\tau)d\tau}{\tau-z}.
\eqno{(3.9)}
$$

Вернемся к решению неоднородной задачи (3.6), предварительно преобразовав
с помощью (3.8) ее к виду:
$$
X^+(\mu)[N^+(\mu)-2U_0]-X^-(\mu)[N^-(\mu)-2U_0]=0,\qquad
\mu>0.
\eqno{(3.10)}
$$

Учитывая поведение всех входящих в краевое условие (3.10)
функций в комплексной плоскости и в бесконечно удаленной точке получаем
общее решение
$$
X(z)[N(z)-2U_0]=C,
\eqno{(3.11)}
$$
где $C$ -- произвольная постоянная.

Согласно (3.11) искомая функция имеет вид
$$
N(z)=2U_0+\dfrac{C}{X(z)}.
\eqno{(3.12)}
$$

Потребуем, чтобы правая часть (3.12) была исчезающей функцией в
бесконечно удаленной точке. Разложим $1/X(z)$
в ряд Лорана в окрестности бесконечно удаленной точки $z=\infty$
$$
\dfrac{1}{X(z)}=\exp\big(-V(z)\big)=1-\dfrac{V_1}{z}+\cdots,
\eqno{(3.13)}
$$
где
$$
V_1=-\dfrac{1}{2\pi i}\int\limits_{0}^{\infty}\ln G(\mu)d\mu.
\eqno{(3.14)}
$$

Таким образом, для того чтобы правая часть равенства (3.12) в бесконечно
удаленной точке имела асимптотическое поведение, как $1/z$,
необходимо на постоянную $C$ наложить условие $C=-2U_0$.
Теперь вспомогательная функция $N(z)$ построена однозначно и имеет вид
$$
N(z)=2U_0\Big[1-\dfrac{1}{X(z)}\Big].
\eqno{(3.15)}
$$

Искомый неизвестный коэффициент непрерывного спектра с помощью (3.9)
находится из формулы Сохоцкого:
$$
2\sqrt{\pi}i\eta a(\eta)=
-2U_0\Big[\dfrac{1}{X^+(\eta)}-\dfrac{1}{X^-(\eta)}\Big]=
4U_0i\dfrac{\sin q(\eta)}{X(\eta)},
$$
где
$$
q(\eta)=-i\dfrac{\ln G(\mu)}{2}=-i\Theta(\mu).
$$
Следовательно,
$$
a(\eta)=\dfrac{2U_0}{\sqrt{\pi}}\dfrac{\sin q(\eta)}{\eta X(\eta)}.
\eqno{(3.16)}
$$

Формула (3.16) дает представление в явном виде коэффициента непрерывного спектра.

На этом этапе доказательство разложения (3.1) (или (3.2)) закончено.

С помощью формулы (3.16) представим разложение (3.3) в явном виде:
$$
\dfrac{h(x_1,\mu)}{2U_0}=\dfrac{1}{\pi}\int\limits_{0}^{\infty}
\exp\Big(-\dfrac{x_1z_0}{\eta}\Big)\Phi(\eta,\mu)
\dfrac{\sin q(\eta)}{\eta X(\eta)}d\eta,
$$
или
$$
\dfrac{h(x_1,\mu)}{2U_0}=\dfrac{1}{\pi}\int\limits_{0}^{\infty}
\exp\Big(-\dfrac{x_1z_0}{\eta}\Big)
\dfrac{\sin q(\eta)d\eta}{X(\eta)(\eta-\mu)}+
$$
$$
+\exp\Big(-\dfrac{x_1z_0}{\mu}+\mu^2)\Big)\dfrac{\lambda(\mu)\sin q(\mu)}{\sqrt{\pi}\mu X(\mu)}\theta_+(\mu).
\eqno{(3.17)}
$$
Равенство (3.17) означает, что искомая функция распределения
построена в явном виде полностью, что и заканчивает аналитическое
решение задачи.

Данная задача с более общими зеркально--диффузными граничными условиями может
быть решена методом, развитым в работах \cite{10} и \cite{11}.

\begin{center}
\item{}\section{Аналитическое решение граничной задачи. Индекс задачи равен единице}
\end{center}

Составим общее решение уравнения (2.11) в виде суммы частного (дискретного) решения, убывающего вдали от стенки, и интеграла по
непрерывному спектру от собственных решений, отвечающих непрерывному спектру:
$$
h(x_1,\mu)=\dfrac{\eta_0a_0}{\sqrt{\pi}(\eta_0-\mu)}\exp\Big(-\dfrac{x_1z_0}
{\eta_0}\Big)+
$$
$$
+\int\limits_{0}^{\infty}
\exp\Big(-\dfrac{x_1z_0}{\eta}\Big)\Phi(\eta,\mu)a(\eta)d\eta.
\eqno{(4.1)}
$$
Здесь $a_0$ -- неизвестный постоянный коэффициент, называемый коэффициентом дискретного спектра, $a(\eta)$ -- неизвестная функция, называемая коэффициентом непрерывного спектра, $\Phi(\eta,\mu)$ -- собственные функции характеристического уравнения,
отвечающие непрерывному спектру и единичной нормировке.

Разложение (4.1) можно представить в явном виде:
$$
h(x_1,\mu)=\dfrac{\eta_0a_0}{\sqrt{\pi}(\eta_0-\mu)}\exp\Big(-\dfrac{x_1z_0}
{\eta_0}\Big)+
$$
$$
+\int\limits_{0}^{\infty}
\exp\Big(-\dfrac{x_1}{\eta}z_0\Big)
\Big[\dfrac{1}{\sqrt{\pi}}\eta P\dfrac{1}{\eta-\mu}+
\exp(\eta^2)\lambda(\eta)\delta(\eta-\mu)\Big]a(\eta)d\eta.
\eqno{(4.2)}
$$
Функция $a(\eta)$ подлежит нахождению из граничных условий (2.12) и (2.13).

Разложение (4.2) можно представить в классическом виде:
$$
h(x_1,\mu)=\dfrac{\eta_0a_0}{\sqrt{\pi}(\eta_0-\mu)}\exp\Big(-\dfrac{x_1z_0}
{\eta_0}\Big)+
$$
$$
+\dfrac{1}{\sqrt{\pi}}\int\limits_{0}^{\infty}
\exp\Big(-\dfrac{x_1z_0}{\eta}\Big)\dfrac{\eta a(\eta)d\eta}{\eta-\mu}+
\exp\Big(-\dfrac{x_1z_0}{\mu}+\mu^2\Big)\lambda(\mu)a(\mu)\theta_+(\mu),
\eqno{(4.3)}
$$
где $\theta_+(\mu)$ -- функция Хэвисайда,
$$
\theta_+(\mu)=\left\{\begin{array}{c}
                       1,\qquad \mu>0, \\
                       0,\qquad \mu<0
                     \end{array}.\right.
$$

Очевидно, что разложение (4.3)
автоматически удовлетворяет граничному условию (2.13) вдали
от стенки. Подставим разложение (4.3) в граничное условие (2.12).
Получаем одностороннее сингулярное интегральное уравнение с ядром Коши
$$
\dfrac{\eta_0a_0}{\sqrt{\pi}(\eta_0-\mu)}+\dfrac{1}{\sqrt{\pi}}
\int\limits_{0}^{\infty}
\dfrac{\eta a(\eta)d\eta}{\eta-\mu}+
\exp(\mu^2)\lambda(\mu)a(\mu)\theta_+(\mu)=2U_0,\; \mu>0.
\eqno{(4.4)}
$$

Введем вспомогательную функцию
$$
N(z)=\dfrac{1}{\sqrt{\pi}}\int\limits_{0}^{\infty}\dfrac{\eta a(\eta)d\eta}
{\eta-\mu}.
\eqno{(4.5)}
$$

Пользуясь формулами Сохоцкого для вспомогательной и дисперсионной функций, от уравнения (4.4) приходим к краевому условию:
$$
\lambda^+(\mu)\bigg[N^+(\mu)-2U_0+
\dfrac{\eta_0a_0}{\sqrt{\pi}(\eta_0-\mu)}\bigg]-
$$
$$
=\lambda^-(\mu)\bigg[N^-(\mu)-2U_0+
\dfrac{\eta_0a_0}{\sqrt{\pi}(\eta_0-\mu)}\bigg]=0, \quad \mu>0.
\eqno{(4.6)}
$$

Рассмотрим соответствующую однородную краевую задачу Римана:
$$
X^+(\mu)=G(\mu)X^-(\mu),\qquad \mu>0,
$$
где
$$
G(\mu)=\dfrac{\lambda^+(\mu)}{\lambda^-(\mu)}.
$$

Решение задачи Римана было рассмотрено в \cite{ALY-2}.
Наличие нулей дисперсионной функции означает,
что приращение $\ln G(\mu)$ на полуоси $[0,+\infty)$ равно $2\pi$.
Поэтому решение задачи Римана дается интегралом типа Коши:
$$
X(z)=\dfrac{1}{z}\exp V(z),
\eqno{(4.7)}
$$
где $V(z)$ понимается как интеграл типа Коши
$$
V(z)=\dfrac{1}{2\pi i}\int\limits_{0}^{\infty}
\dfrac{[\ln G(\tau)-2\pi i]d\tau}{\tau-z}.
\eqno{(4.8)}
$$

Вернемся к решению неоднородной задачи (4.6), предварительно преобразовав
с помощью (4.7) ее к виду:
$$
X^+(\mu)\bigg[N^+(\mu)-2U_0+\dfrac{\eta_0a_0}{\sqrt{\pi}(\eta_0-\mu)}\bigg]-
$$
$$
-X^-(\mu)\bigg[N^-(\mu)-2U_0+\dfrac{\eta_0a_0}{\sqrt{\pi}(\eta_0-\mu)}\bigg]=0,
\quad \mu>0.
\eqno{(4.9)}
$$

Учитывая поведение всех входящих в краевое условие (4.9)
функций в комплексной плоскости и в бесконечно удаленной точке получаем
общее решение
$$
X(z)\bigg[N(z)-2U_0+\dfrac{\eta_0a_0}{\sqrt{\pi}(\eta_0-\mu)}\bigg]=
\dfrac{C}{z-\eta_0},
\eqno{(4.10)}
$$
где $C$ -- произвольная постоянная.

Согласно (4.10) искомая функция имеет вид
$$
N(z)=2U_0+\dfrac{\eta_0a_0}{\sqrt{\pi}(z-\eta_0)}+\dfrac{C}{X(z)(z-\eta_0)}.
\eqno{(4.11)}
$$

Полюс в точке $z=\eta_0$ у решения (4.11) устраним условием:
$$
\dfrac{C}{X(\eta_0)}+\dfrac{\eta_0a_0}{\sqrt{\pi}}=0,
$$
откуда находим связь коэффициентов $C$ и $a_0$:
$$
C=-\dfrac{\eta_0a_0}{\sqrt{\pi}}X(\eta_0).
\eqno{(4.12)}
$$

Потребуем, чтобы правая часть (4.12) была исчезающей функцией в
бесконечно удаленной точке. На этом пути получаем:
$$
C=-2U_0.
$$
Следовательно, из условия (4.12) находим:
$$
a_0=2U_0\sqrt{\pi}\dfrac{1}{\eta_0X(\eta_0)}=2U_0\sqrt{\pi}\exp(-V(\eta_0)).
\eqno{(4.13)}
$$

С помощью коэффициентов $C$ и $a_0$, определяемых равенствами (4.12) и (4.13), преобразуем решение (4.11) к следующему виду:
$$
N(z)=2U_0+\dfrac{2u_0}{X(\eta_0)(z-\eta_0)}-\dfrac{2U_0}{X(z)(z-z_0)}.
\eqno{(4.14)}
$$

Искомый неизвестный коэффициент непрерывного спектра с помощью (4.14)
находится из формулы Сохоцкого:

$$
2\sqrt{\pi}i\eta a(\eta)=
-\dfrac{2U_0}{\eta-\eta_0}\Big[\dfrac{1}{X^+(\eta)}-\dfrac{1}{X^-(\eta)}\Big]=
4U_0i\dfrac{\sin q(\eta)}{X(\eta)(\eta-\eta_0)},
$$
где
$$
q(\eta)\equiv i\Theta(\eta) \equiv -\dfrac{i}{2}(\ln G(\mu)-2\pi i).
$$
Следовательно,
$$
a(\eta)=\dfrac{2U_0}{\sqrt{\pi}}\dfrac{\sin q(\eta)}{\eta X(\eta)(\eta-\eta_0)}.
\eqno{(4.15)}
$$

Формула (4.15) дает представление в явном виде коэффициента непрерывного спектра.

На этом этапе доказательство разложения (4.1) (или (4.2)) закончено.

С помощью формулы (4.15) представим разложение (4.3) в явном виде:
$$
\dfrac{h(x_1,\mu)}{2U_0}=\dfrac{1}{X(\eta_0)(\eta_0-\mu)}\exp\Big(-
\dfrac{x_1z_0}{\eta_0}\Big)+
$$
$$
+\dfrac{1}{\pi}\int\limits_{0}^{\infty}
\exp\Big(-\dfrac{x_1z_0}{\eta}\Big)\dfrac{\sin q(\eta)d\eta}{X(\eta)(\eta-\eta_0)(\eta-\mu)}+
$$
$$
+\exp\Big(-\dfrac{x_1z_0}{\mu}+\mu^2\Big)
\dfrac{\lambda(\mu)\sin q(\mu)}{\sqrt{\pi}\mu X(\mu)(\mu-\eta_0)}\theta_+(\mu).
\eqno{(4.16)}
$$
Равенство (4.16) означает, что искомая функция распределения
построена в явном виде полностью, что и заканчивает аналитическое
решение задачи.

Данная задача с более общими зеркально--диффузными граничными условиями может быть решена методом, развитым в работах \cite{10} и \cite{11}.

\begin{center}
\item{}\section{Скорость разреженного газа в полупространстве и непосредственно у колеблющейся плоскости}
\end{center}

Начнем со случая, когда индекс задачи равен нулю. В этом случае частота колебаний пластины $\omega_1\in (\omega_1^*,+\infty)$.

В п. 2 было найдено выражение (2.10) для безразмерной массовой скорости.
Упростим это выражение. Воспользуемся разложением (3.1).
Подставим (3.1) в (2.10) и поменяем порядок интегрирования. Затем, используя нормировочное соотношение $n(\eta)\equiv 1$, приходим к равенству:
$$
U_y(x_1,t_1)=\dfrac{1}{2\sqrt{\pi}}\Re\bigg\{e^{-i\omega_1t_1}z_0
\int\limits_{0}^{\infty}\exp\Big(-\dfrac{x_1}{\eta}z_0\Big)a(\eta)d\eta\bigg\}.
\eqno{(5.1)}
$$
Теперь воспользуемся формулой (3.13) для коэффициента непрерывного спектра. В результате получим, что массовая скорость газа
в полупространстве равна:
$$
U_y(x_1,t_1)=\dfrac{U_0}{{\pi}}
\Re\bigg\{e^{-i\omega_1t_1}z_0
\int\limits_{0}^{\infty}\exp\Big(-\dfrac{x_1}{\eta}z_0\Big)
\dfrac{\sin q(\eta)}{\eta X(\eta)}d\eta\bigg\}.
\eqno{(5.2)}
$$

Вычислим значение массовой скорости непосредственно вблизи у стенки. Из формулы (5.2) получаем, что
$$
U_y(0,t_1)=\dfrac{U_0}{{\pi}}\Re\bigg\{e^{-i\omega_1t_1}z_0
\int\limits_{0}^{\infty}\dfrac{\sin q(\eta)}{\eta X(\eta)}d\eta\bigg\}.
\eqno{(5.3)}
$$

Для вычисление интеграла из (5.3) воспользуемся интегральным представлением (см. \cite{ALY-2}):
$$
\dfrac{1}{X(z)}-1=-\dfrac{1}{\pi}\int\limits_{0}^{\infty}
\dfrac{\sin q(\eta)d\eta}{X(\eta)(\eta-z)}.
\eqno{(5.4)}
$$

Из (5.4) видно, что
$$
\dfrac{1}{X(0)}-1=-\dfrac{1}{\pi}\int\limits_{0}^{\infty}
\dfrac{\sin q(\eta)d\eta}{\eta X(\eta)}.
$$
Отсюда находим, что
$$
\dfrac{1}{\pi}\int\limits_{0}^{\infty}\dfrac{\sin q(\eta)d\eta}{\eta X(\eta)}=1-\dfrac{1}{X(0)}.
\eqno{(5.5)}
$$

Следовательно, массовая скорость в полупространстве вычисляется по формуле:
$$
U_y(0,t_1)=U_0\Re\bigg\{e^{-i\omega_1t_1}z_0\Big(1-\dfrac{1}{X(0)}\Big)\bigg\}.
\eqno{(5.6)}
$$

Для нахождения величины факторизующей функции в нуле воспользуемся теперь формулой факторизации дисперсионной функции \cite{ALY-2}:
$$
\lambda(z)=\lambda_\infty X(z)X(-z),
\eqno{(5.7)}
$$
где
$$
\lambda_\infty=\lambda(\infty)=-i\omega_1.
$$

Замечая, что $\lambda(0)=1-i\omega_1$, из (5.7) находим:
$$
X^2(0)=\dfrac{\lambda(0)}{\lambda_\infty}=\dfrac{1-i\omega_1}{-i\omega_1}=
1+\dfrac{i}{\omega_1}=\dfrac{\omega_1+i}{\omega_1}=\dfrac{\omega+i \nu}{\omega},
$$
откуда
$$
X(0)=\sqrt{1+\dfrac{i}{\omega_1}}=\dfrac{\sqrt{\omega_1+i}}{\sqrt{\omega_1}}=
\sqrt{\dfrac{\omega+i \nu}{\omega}}.
\eqno{(5.8)}
$$

Следовательно, согласно (5.6)--(5.8) находим значение амплитуды скорости газа у стенки:
$$
U_y(0,t_1)=U_0\Re\bigg\{e^{-i\omega_1t_1}(1-i\omega_1)
\dfrac{\sqrt{\omega_1+i}-\sqrt{\omega_1}}{\sqrt{\omega_1+i}}\bigg\}.
\eqno{(5.9)}
$$

Значение размерной скорости непосредственно у стенки дается выражением:
$$
u_y(0,t)=v_TU_0\Re\bigg\{e^{-i\omega t}(\nu-i\omega)
\dfrac{\sqrt{\omega+i \nu}-\sqrt{\omega}}{\nu\sqrt{\omega+i \nu}}\bigg\}.
$$

Обозначим
$$
W=(1-i\omega_1)\dfrac{\sqrt{\omega_1+i}-\sqrt{\omega_1}}{\sqrt{\omega_1+i}}=
(\nu-i\omega)
\dfrac{\sqrt{\omega+i \nu}-\sqrt{\omega}}{\nu\sqrt{\omega+i \nu}}
\eqno{(5.10)}
$$
и перепишем (5.9) в виде
$$
U_y(0,t_1)=U_0 \Re\{e^{-i\omega_1t_1}W(\omega_1)\},
\eqno{(5.9')}
$$
откуда, полагая, что $W=|W|e^{i\varphi}$, запишем
$$
U_y(0,t_1)=U_0|W|\cos (\omega_1t_1-\varphi),
\eqno{(5.10')}
$$
или, в размерном виде
$$
U_y(0,t_1)=v_TU_0|W|\cos (\omega t-\varphi),
\eqno{(5.10'')}
$$
где $|W|$ -- безразмерная амплитуда скорости газа (см. рис. 1), а $\varphi=\arg W$ -- сдвиг фазы скорости (см. рис. 2).

Теперь рассмотрим случай, когда индекс задачи равен единице, т.е. когда параметр $\omega_1\in [0,\omega_1^*)$. Подставим решение (4.1) в формулу (2.10) для скорости газа. В результате получаем следующее выражение
$$
U_y(x_1,t_1)=\dfrac{U_0}{\sqrt{\pi}}
\Re\Bigg\{\dfrac{\exp\Big(-\dfrac{x_1z_0}{\eta_0}-
i\omega_1t_1\Big)}{X(\eta_0)}\dfrac{1}{\sqrt{\pi}}\int\limits_{-\infty}^{\infty}
\dfrac{e^{-\tau^2}d\tau}{\eta_0-\tau}+
$$
$$
+\dfrac{e^{-i\omega_1t_1}}{\sqrt{\pi}}
\int\limits_{0}^{\infty}\dfrac{\exp\Big(-\dfrac{x_1z_0}{\eta}\Big)
\sin q(\eta)d\eta}{\eta X(\eta)(\eta-\eta_0)}\int\limits_{-\infty}^{\infty}
e^{-\tau^2}\Phi(\eta,\mu)d\tau\Bigg\}.
$$

Заметим, что из уравнения $\lambda(\eta_0)=0$ вытекает следующее равенство
$$
\dfrac{1}{\sqrt{\pi}}\int\limits_{-\infty}^{\infty}\dfrac{e^{-\tau^2}d\tau}
{\eta_0-\tau}=\dfrac{z_0}{\eta_0},
$$
а условие нормировки собственных функций дает выражение
$$
\int\limits_{-\infty}^{\infty}e^{-\mu^2}\Phi(\eta,\mu)d\mu \equiv z_0.
$$

С помощью двух последних равенств выражение скорости газа упрощается:
$$
U_y(x_1,t_1)=U_0\Re\Bigg\{e^{-i\omega_1t_1}z_0
\bigg[\dfrac{e^{-x_1z_0/\eta_0}}{\eta_0X(\eta_0)}+\dfrac{1}{\pi}
\int\limits_{0}^{\infty}\dfrac{e^{-x_1z_0/\eta}\sin q(\eta)d\eta)}{\eta X(\eta)(\eta-\eta_0)}\bigg]\Bigg\}.
\eqno{(5.11)}
$$

Эта формула дает выражение для скорости газа над колеблющейся поверхностью в полупространстве $x_1>0$. Согласно (5.11) при $x_1=0$ найдем скорость газа непосредственно у стенки:
$$
U_y(0,t_1)=U_0\Re\Bigg\{e^{-i\omega_1t_1}z_0
\bigg[\dfrac{1}{\eta_0X(\eta_0)}+\dfrac{1}{\pi}
\int\limits_{0}^{\infty}\dfrac{\sin q(\eta)d\eta)}{\eta X(\eta)(\eta-\eta_0)}\bigg]\Bigg\}.
\eqno{(5.12)}
$$

Для вычисления интеграла из (5.12) воспользуемся интегральным представлением
$$
\dfrac{1}{X(z)}-z+V_1=-\dfrac{1}{\pi}\int\limits_{0}^{\infty}
\dfrac{\sin q(\eta)d\eta}{X(\eta)(\eta-z)}, \qquad
V_1=-\dfrac{1}{2\pi i}\int\limits_{0}^{\infty}[\ln G(\tau)-2\pi i]d\tau,
$$
и разложением на элементарные дроби
$$
\dfrac{1}{\eta(\eta-\eta_0)}=\dfrac{1}{\eta_0}\Big(\dfrac{1}{\eta-\eta_0}-
\dfrac{1}{\eta}\Big).
$$
Теперь этот интеграл равен:
$$
\dfrac{1}{\pi}\int\limits_{0}^{\infty}\dfrac{\sin q(\eta)d\eta)}{\eta X(\eta)(\eta-\eta_0)}=\dfrac{1}{\eta_0}\Bigg[\int\limits_{0}^{\infty}\dfrac{\sin q(\eta)d\eta)}{X(\eta)(\eta-\eta_0)}-\int\limits_{0}^{\infty}\dfrac{\sin q(\eta)d\eta)}{\eta X(\eta)}
\Bigg]=
$$
$$
=-1+\dfrac{1}{\eta_0X(\eta_0)}-\dfrac{1}{\eta_0X(0)}.
$$

С помощью этого равенства выражение для скорости газа непосредственно у стенки упрощается:
$$
U_y(0,t_1)=U_0\Re\Big[e^{-i\omega_1t_1}z_0\Big(1+\dfrac{1}{\eta_0X(0)}\Big)\Big].
\eqno{(5.13)}
$$

Для вычисления величины $X(0)$ из (5.13) воспользуемся формулой факторизации дисперсионной функции:
$$
\lambda(z)=i\omega_1(z^2-\eta_0^2)X(z)X(-z),
\eqno{(5.14)}
$$
которая была доказана в \cite{ALY-2}.

Из формулы (5.14) при $z=0$ находим, что
$$
\lambda(0)=-i\omega_1\eta_0^2X^2(0),
$$
откуда, учитывая, что $\lambda(0)=z_0$, получаем:
$$
X(0)=\sqrt{\dfrac{iz_0}{\omega_1\eta_0^2}}.
$$

Следовательно, согласно (5.13) для безразмерной скорости газа получаем
$$
U_y(0,t_1)=U_0|W|\cos(\omega_1t_1-\varphi),
$$
где
$$
W=z_0+(1-i)\sqrt{\dfrac{\omega_1z_0}{2}},
\eqno{(5.15)}
$$
$|W|$ -- амплитуда скорости, а $\varphi=\arg W$ -- сдвиг фазы скорости.

Итак, выражение для размерной скорости при малых частотах непосредственно у стенки таково:
$$
u_y(0,t_1)=v_TU_0\Re\Big[e^{-i\omega_1t_1}z_0
\Big(1+\sqrt{\dfrac{\omega_1}{z_0}}\cdot e^{-i\dfrac{\pi}{4}}\Big)\Big],
$$
или
$$
u_y(0,t_1)=v_TU_0\Re\bigg[e^{-i\omega_1t_1}
\Big(z_0+(1-i)\sqrt{\dfrac{\omega_1 z_0}{2}}\Big)\bigg]
\eqno{(5.16)}
$$
Отметим, что при малых $\omega_1$ $\eta_0^2\approx i/2\omega_1$. Следовательно,
$$
X(0)=\sqrt{2z_0}=\sqrt{2(1-i\omega_1)}.
$$
\begin{figure}[t]
\begin{center}
\includegraphics[width=15.0cm, height=10cm]{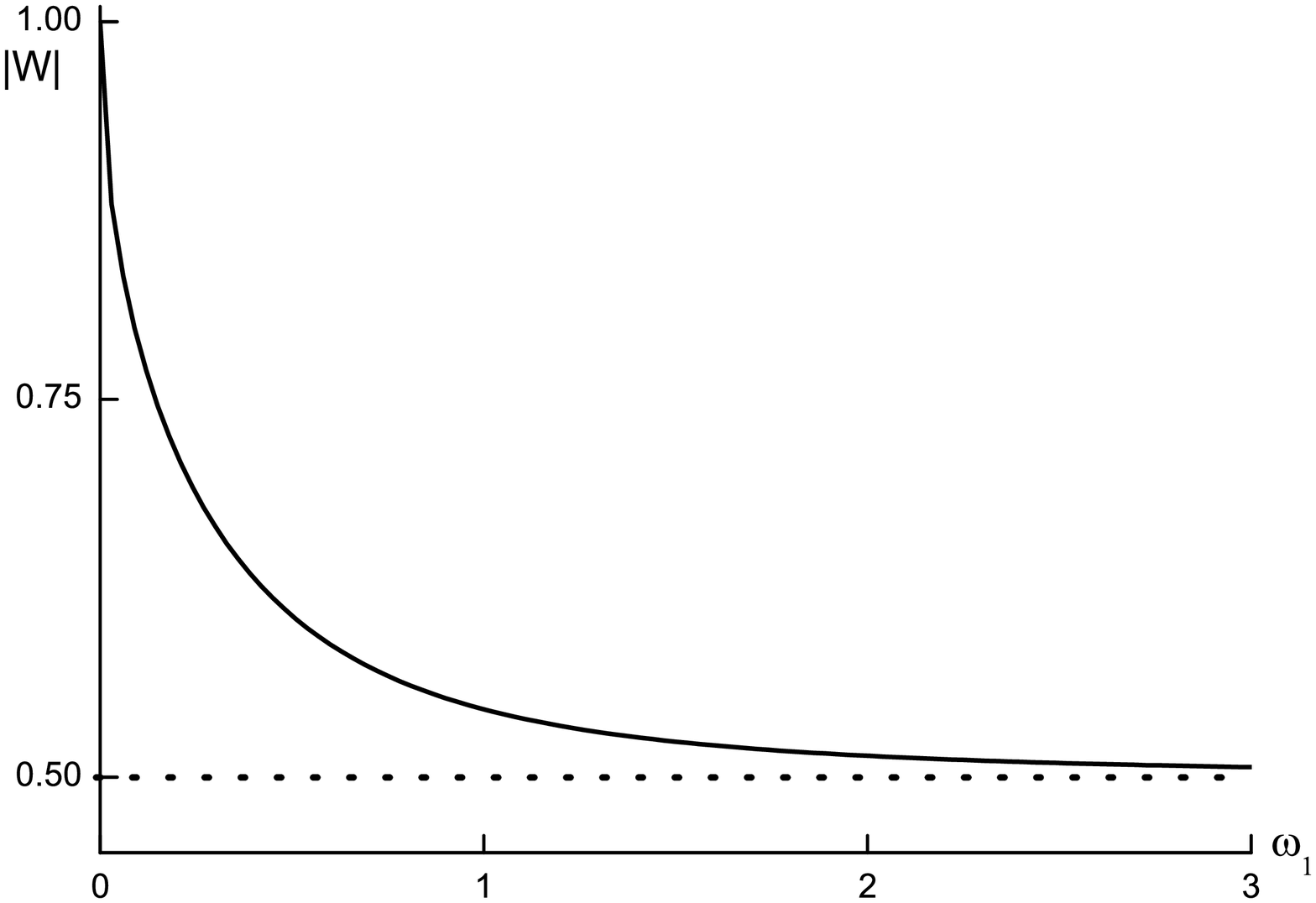}
\end{center}
\begin{center}
{{Рис. 1. Зависимость величины скорости газа непосредственно у стенки
от частоты колебаний ограничивающей газ плоскости. Эта зависимость при
$0\leqslant \omega_1 \leqslant \omega_1^*$ построена по формуле (5.13), а при $\omega \geqslant \omega_1^*$ -- по формуле (5.10).}}
\end{center}
\end{figure}
\begin{figure}[t]
\begin{center}
\includegraphics[width=15.0cm, height=10cm]{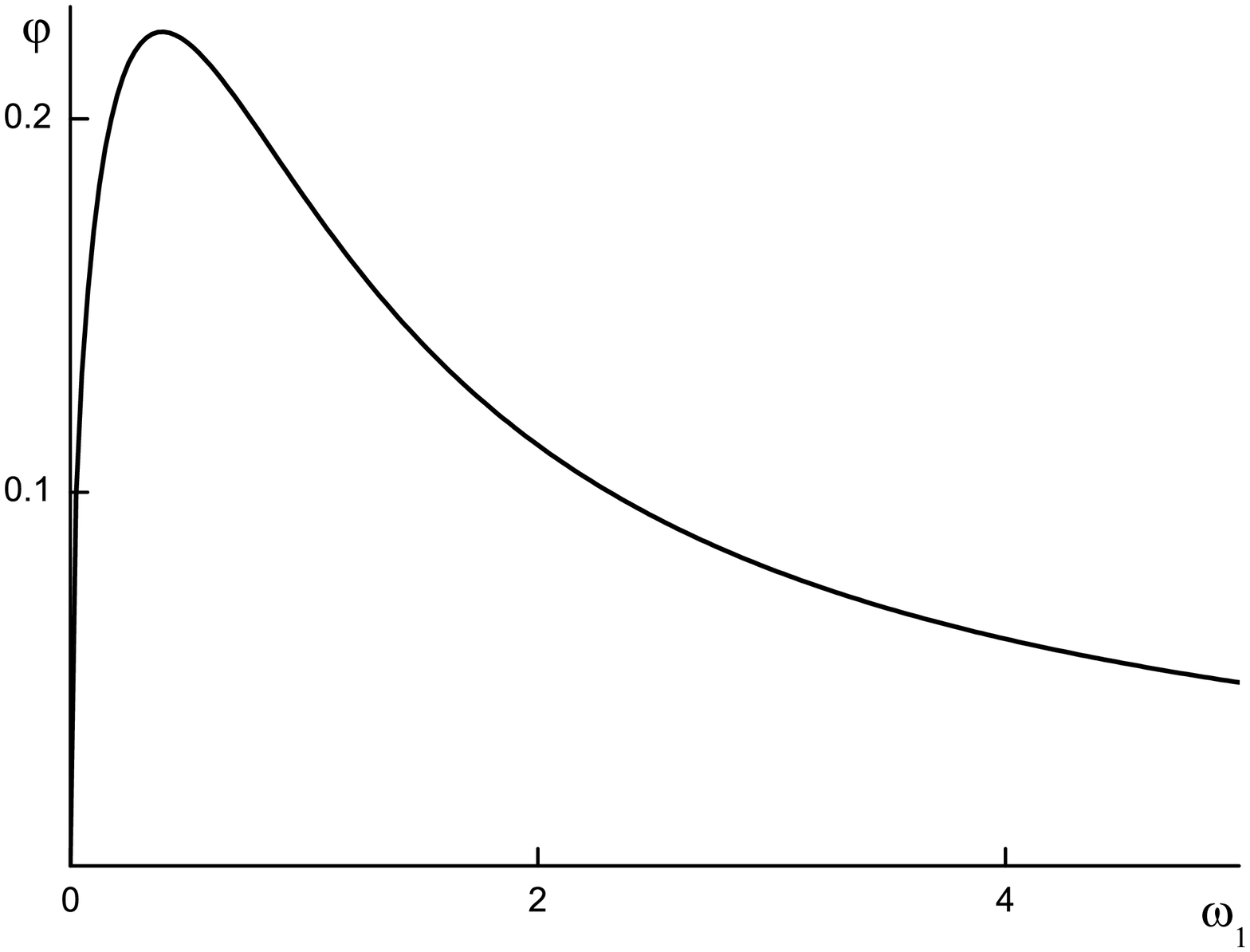}
\end{center}
\begin{center}
{{Рис. 2. Зависимость величины сдвига фазы скорости газа непосредственно у стенки от частоты колебаний ограничивающей газ плоскости. Эта зависимость при
$0\leqslant \omega_1 \leqslant \omega_1^*$ построена по формуле (5.15), а при $\omega \geqslant \omega_1^*$ -- по формуле (5.10).}}
\end{center}
\end{figure}

\begin{center}
\item{}\section{О гидродинамическом характере решения}
\end{center}

В этом п. покажем, что при малых $\omega_1$ решение (5.11) переходит в решение, приведенное в \cite{LandauG}:
$$
v=u_0e^{-x/\delta}e^{i(x/\delta-\omega t)}.
\eqno{(6.1)}
$$

Здесь
$$
\delta=\sqrt{\dfrac{2\nu_k}{\omega}},
$$
$\nu_k$ -- кинематическая вязкость газа.

Формула (6.1) выведена для случая сплошной среды в случае, когда ограничивающая среду плоскость совершает гармонические колебания по закону $U_s(t)=U_0e^{-i\omega t}$. При таком законе колебаний плоскости в нашей задаче мы приходим к формуле (5.11), в которой отброшена операция $\Re$ --- операция взятия действительной части:
$$
U_y(x_1,t_1)=U_0e^{-i\omega_1t_1}z_0
\bigg[\dfrac{e^{-x_1z_0/\eta_0}}{\eta_0X(\eta_0)}-\dfrac{1}{\pi}
\int\limits_{0}^{\infty}\dfrac{e^{-x_1z_0/\eta}\sin q(\eta)d\eta)}{\eta X(\eta)(\eta-\eta_0)}\bigg].
\eqno{(6.2)}
$$

При малых $\omega_1$ нуль дисперсионной функции $\eta_0\to \infty$,
следовательно, интеграл по непрерывному спектру является исчезающе малым. Далее заметим, что
$$
\eta_0X(\eta_0)=e^{V(\eta_0)},
$$
а при больших значений $|\eta_0|$ интеграл $V(\eta_0)$ исчезает при малых
$\omega_1$. Значит, при $\omega_1\to 0$ для скорости газа получаем выражение
$$
u_y(x_1,t_1)=u_0e^{-i\omega_1t_1}e^{-x_1/\eta_0}.
\eqno{(6.3)}
$$

Здесь $\omega_1t_1=\omega t$, $x_1=x/l, l=\tau /\sqrt{\beta}$.
При малых $\omega_1$ для нуля дисперсионной функции справедливо представление
$$
\eta_0=\dfrac{1+i}{2\sqrt{\omega_1}}.
$$
Следовательно, выражение (6.3) преобразуется далее следующим образом:
$$
u_y(x,t)=u_0e^{-i\omega t}e^{-x/l\eta_0}.
$$
Замечая, что $\dfrac{\tau}{\beta}=2\nu_k$, далее получаем:
$$
\dfrac{1}{l\eta_0}=\dfrac{\sqrt{\beta}2\sqrt{\omega\tau}}{\tau(1+i)}=
\dfrac{1-i}{\sqrt{\dfrac{\tau}{\omega \beta}}}=\dfrac{1-i}{\sqrt{\dfrac{2\nu_k}{\omega}}}=\dfrac{1-i}{\delta}.
$$
Это означает, что
$$
u_y(x,t)=u_0e^{-i\omega t}e^{-(1-i)x/\delta},
$$
что в точности совпадает с выражением (24,5) из \cite{LandauG}.

\begin{center}
\item{}\section{Сила трения, действующая со стороны газа на колеблющуюся границу, и диссипация энергии}
\end{center}

Сила трения, приходящаяся на единицу площади, действующая со стороны газа на пластину, вычисляется по формуле
$$
P_{xy}(x,t)\Big|_{x=0}=m\int v_xv_yf\Big|_{x=0}d^3v.
\eqno{(6.1)}
$$

Функция распределения построена в предыдущих п.п. и имеет вид:
$$
f=f_M(C)\Big[1+C_y\Re\{e^{-i\omega_1t_1}h(x_1,C_x)\}\Big].
$$

Теперь компонента тензора вязких напряжений вычисляется по формуле:
$$
P_{xy}(x_1,t_1)=\dfrac{mn}{\pi^{3/2}\beta}\int \exp(-C^2)C_xC_y^2
\Re\bigg[e^{-i \omega_1t_1}h(x_1,C_x)\bigg]d^3C,
$$
или, после интегрирования по $C_y$ и $C_z$,
$$
P_{xy}(x_1,t_1)=\dfrac{mn}{2\beta \sqrt{\pi}}\int\limits_{-\infty}^{\infty}e^{-\mu^2}
\Re\bigg[e^{-i \omega_1t_1}\mu h(x_1,\mu)\bigg]d\mu.
$$
где функция $h(x_1,\mu)$ определяется равенством (3.1) в случае нулевого
индекса задачи, и разложением (4.1) в случае единичного индекса.

Замечая, что
$$
\dfrac{mn}{2\beta}=nkT=p,
$$
перепишем предыдущую формулу в виде
$$
P_{xy}(x_1,t_1)=\dfrac{p}{\sqrt{\pi}}\int\limits_{-\infty}^{\infty}e^{-\mu^2}
\Re\bigg[e^{-i \omega_1t_1}\mu h(x_1,\mu)\bigg]d\mu.
\eqno{(6.2)}
$$

Начнем со случая, когда индекс задачи равен нулю. В этом случае частота колебаний пластины $\omega_1\in (\omega_1^*,+\infty)$.

Подставляя разложение (3.1) в (6.2), получаем:
$$
P_{xy}(x_1,t_1)=\dfrac{nkT}{\sqrt{\pi}}\Re\bigg\{e^{-i \omega_1t_1 } \int\limits_{0}^{\infty}e^{-{x_1z_0}/{\eta}}
a(\eta)d\eta \int\limits_{-\infty}^{\infty}e^{-\mu^2}\mu \Phi(\eta,\mu)d\mu\bigg\}.
\eqno{(6.3)}
$$

Обозначим
$$
n_1(\eta)=\int\limits_{-\infty}^{\infty}\exp(-\mu^2)\mu\Phi(\eta,\mu)d\mu.
$$

Эту величину вычислим, используя характеристическое уравнение. Умножим характеристическое уравнение на $\exp(-\mu^2)$ и проинтегрируем по $\mu$. Получаем уравнение
$$
\eta z_0 n(\eta)-n_1(\eta)=\eta n(\eta),
$$
откуда
$$
n_1(\eta)=-i\omega_1 \eta, \quad\text{ибо}\quad n(\eta)\equiv 1.
$$

Согласно (6.3) компонента тензора вязких напряжений равна
$$
P_{xy}(x_1,t_1)=-\dfrac{p}{\sqrt{\pi}}\Re\bigg\{e^{-i \omega_1t_1}i \omega_1 \int\limits_{0}^{\infty}e^{-x_1z_0/\eta}\eta a(\eta)d\eta\bigg\}.
\eqno{(6.4)}
$$

Таким образом, сила, приходящаяся на единицу площади пластины, равна
$$
F_s(t_1)=P_{x,y}(0,t_1)=-\dfrac{p}{\sqrt{\pi}}\Re\bigg\{e^{-i \omega_1t_1}i \omega_1 \int\limits_{0}^{\infty}\eta a(\eta)d\eta\bigg\}.
$$

Здесь $p=nkT$ -- величина давления в газе.

Пользуясь формулой (4.15) для коэффициента непрерывного спектра, окончательно получаем формулу для вычисления силы, действующей на единицу
площади колеблющейся пластины, ограничивающей газ:
$$
F_s(t_1)=-\dfrac{2pU_0}{\pi}\Re\bigg\{e^{-i\omega_1t_1}i \omega_1
\int\limits_{0}^{\infty}\dfrac{\sin q(\eta)d\eta}{X(\eta)}\bigg\}.
$$

Из рассуждений п. 4 следует, что
$$
\dfrac{1}{\pi}\int\limits_{0}^{\infty}
\dfrac{\sin q(\eta)d\eta}{X(\eta)}=-V_1,
$$
где
$$
V_1=-\dfrac{1}{2\pi i}\int\limits_{0}^{\infty}\ln G(\eta)d\eta.
$$
Окончательно, выражение для силы таково:
$$
F_s(t_1)=2pU_0\omega_1\Re\{i e^{-i \omega_1t_1}V_1\}.
\eqno{(6.5)}
$$
Выделяя действительную часть в выражении (6.5), получаем:
$$
F_s(t_1)=2pU_0\omega_1|V_1|\sin(\omega_1t_1-\varphi),
\eqno{(6.6)}
$$
где $\varphi$ -- сдвиг фазы, $\varphi=\arg V_1$.

Величина сдвига фазы может быть определена из соотношений
$$
\sin\varphi=\dfrac{\Im V_1}{|V_1|},\quad
\cos \varphi=\dfrac{\Re V_1}{|V_1|}.
\eqno{(6.7)}
$$

Теперь рассмотрим случай, когда индекс задачи равен единице, т.е. когда параметр $\omega_1\in [0,\omega_1^*)$.

Воспользуемся формулой (4.1) для функции $h(x_1,t_1)$. Получаем, что
$$
P_{xy}(x_1,t_1)=\dfrac{p}{\sqrt{\pi}}\int\limits_{-\infty}^{\infty}
e^{-\mu^2}\Re\Bigg\{e^{-i\omega_1t_1}\mu\Big[\dfrac{\eta_0a_0e^{-x_1z_0/\eta_0}}
{\sqrt{\pi}(\eta_0-\mu)}\Big]+
$$
$$
+\int\limits_{0}^{\infty}e^{-x_1z_0/\eta}\Phi(\eta,\mu)a(\eta)d\eta\Bigg\},
$$
или,
$$
P_{xy}(x_1,t_1)=\dfrac{p}{\sqrt{\pi}}\Re\Bigg\{e^{-i\omega_1t_1}\Bigg[
-\eta_0a_0e^{-x_1z_0/\eta_0}\dfrac{1}{\sqrt{\pi}}\int\limits_{-\infty}^{\infty}
\dfrac{e^{-\mu^2}\mu d\mu}{\mu-\eta_0}+
$$
$$
+\int\limits_{-\infty}^{\infty}e^{-\mu^2}\mu d\mu \int\limits_{0}^{\infty}e^{-x_1z_0/\eta}\Phi(\eta,\mu)a(\eta)d\eta\Bigg]\Bigg\}.
\eqno{(6.8)}
$$

Вычислим два интеграла из (6.8).
Первый интеграл вычисляется на основании определения нуля дисперсионной функции: $\lambda(\eta_0)=0$. Отсюда следует, что
$$
\dfrac{1}{\sqrt{\pi}}\int\limits_{-\infty}^{\infty}\dfrac{e^{-\mu^2}\mu d\mu}{\mu-\eta_0}=\lambda_0(\eta_0)=i\omega_1.
$$

В повторном интеграле из (6.8) поменяем местами порядок интегрирования и вычислим внутренний интеграл:
$$
\int\limits_{-\infty}^{\infty}e^{-\mu^2}\mu \Phi(\eta,\mu)d\mu
\equiv n_1(\mu)=-i\omega_1 \eta.
$$
С помощью двух последних равенств выражение (6.8) упрощается:
$$
P_{xy}(x_1,t_1)=-\dfrac{p\omega_1}{\sqrt{\pi}}\Re\bigg[ie^{-i\omega_1t_1}
\Big[\eta_0a_0e^{-x_1z_0/\eta_0}+\int\limits_{0}^{\infty}
e^{-x_1z_0/\eta}\eta a(\eta)\Big]\bigg].
$$
Теперь воспользуемся выражениями для коэффициентов дискретного и непрерывного спектров. В результате получаем следующее выражение:
$$
P_{xy}(x_1,t_1)=-2U_0p\omega_1\Re\bigg[ie^{-i\omega_1t_1}
\Big[\dfrac{e^{-x_1z_0/\eta_0}}{X(\eta_0)}+\dfrac{1}{\pi}
\int\limits_{0}^{\infty}\dfrac{e^{-x_1z_0/\eta}\sin q(\eta)d\eta}{X(\eta)(\eta-\eta_0)}\Big]\bigg].
\eqno{(6.9)}
$$

На основании (6.9) находим значение тензора вязких напряжений на границе полупространства:
$$
P_{xy}(0,t_1)=-2U_0p\omega_1\Re\Bigg\{ie^{-i\omega_1t_1}
\bigg[\dfrac{1}{X(\eta_0)}+
\dfrac{1}{\pi}\int\limits_{0}^{\infty}\dfrac{\sin q(\eta)d\eta}
{X(\eta)(\eta-\eta_0)}\bigg]\Bigg\}.
$$

Воспользуемся интегральным представлением из \cite{ALY-2}
$$
\dfrac{1}{\pi}\int\limits_{0}^{\infty}\dfrac{\sin q(\eta)d\eta}
{X(\eta)(\eta-\eta_0)}=-\dfrac{1}{X(\eta_0)}+\eta_0-V_1.
$$

\begin{figure}[t]
\begin{center}
\includegraphics[width=15.0cm, height=9cm]{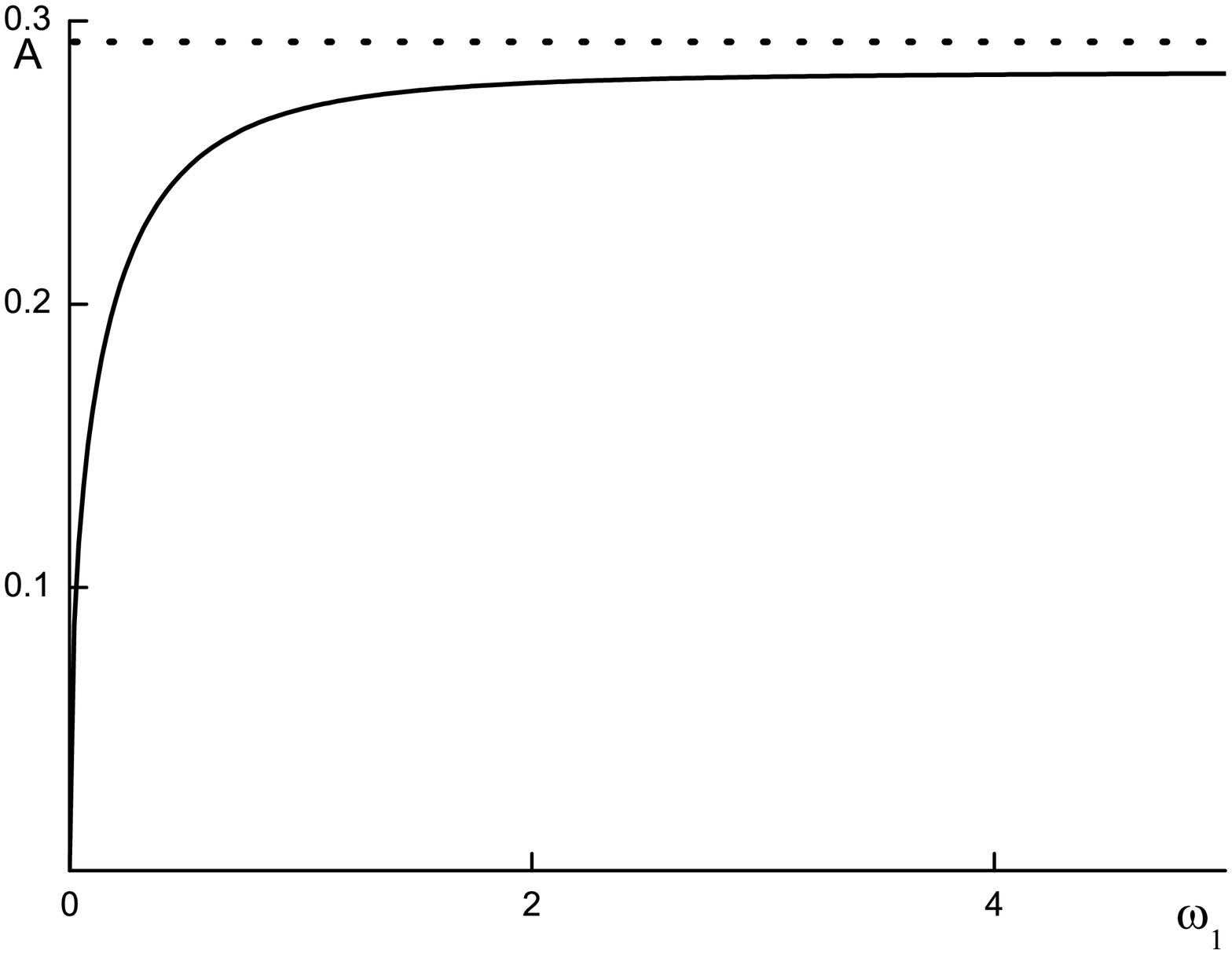}
\end{center}
\begin{center}
{{Рис. 3. Зависимость величины амплитуды силы трения от частоты колебаний ограничивающей газ плоскости. Эта зависимость построена по формуле (6.10a).}}
\includegraphics[width=15.0cm, height=9cm]{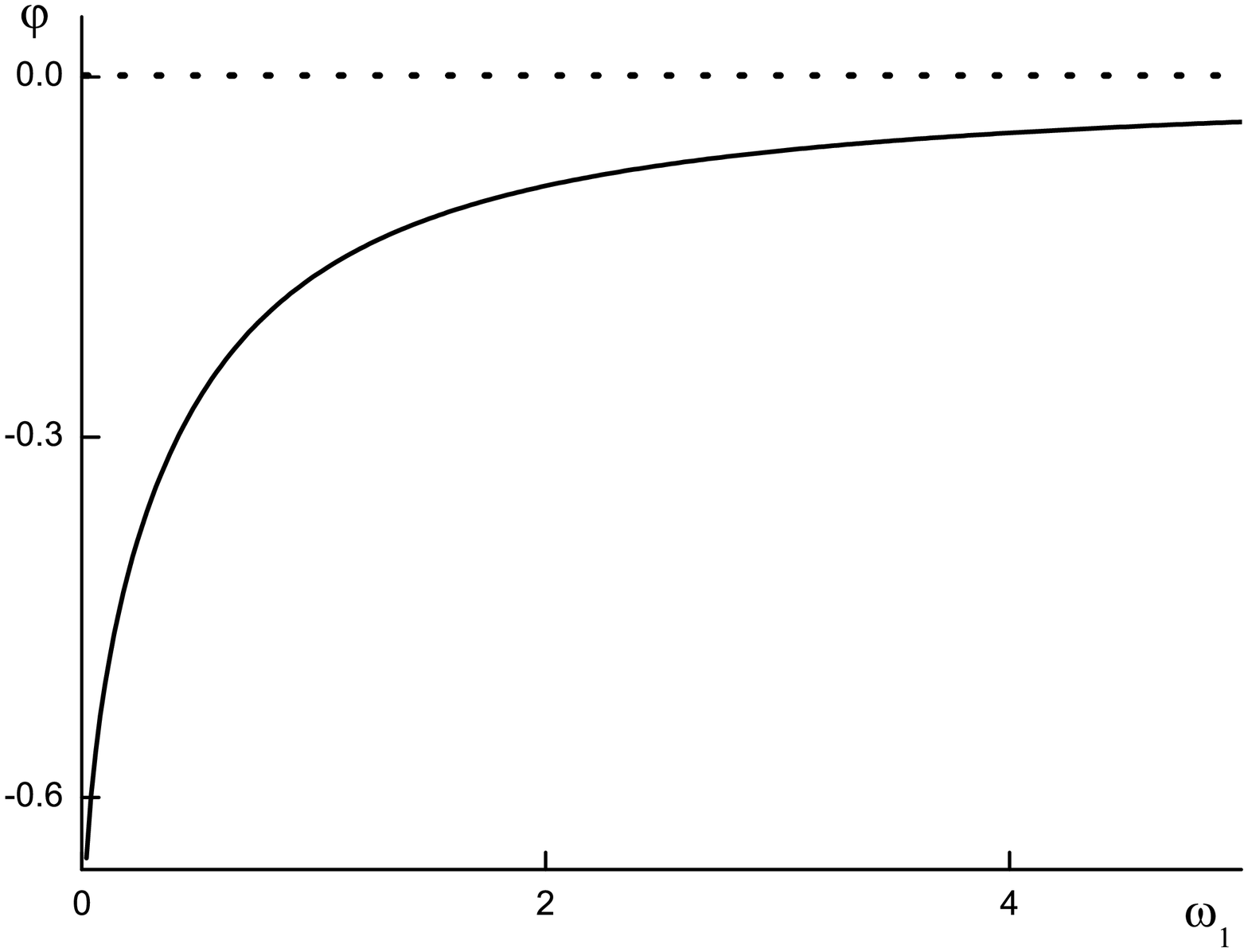}
\end{center}
\begin{center}
{{Рис. 4. Зависимость сдвига фазы силы трения от частоты колебаний ограничивающей газ плоскости. Эта зависимость построена по формуле (6.10b).}}
\end{center}
\end{figure}
\clearpage

Отсюда получаем, что
$$
P_{xy}(0,t_1)=-2U_0p\omega_1\Re\{e^{-i\omega_1t_1}i(\eta_0-V_1)\}.
$$

В результате получаем, что на границе полупространства сила трения газа, действующая на границу, равна:
$$
F_s(t_1)=P_{xy}(0,t_1)=\Re(F_0e^{-i\omega_1t_1}),
\eqno{(6.10)}
$$
где
$$
F_0=-2U_0p\omega_1\big[i(\eta_0-V_1)\big].
$$

График амплитуды силы трения (см. (рис. 3)) построим по формуле
$$
A=\left\{\begin{array}{c}
           \omega_1|\eta_0-V_1|,\quad 0 \leqslant \omega_1 \leqslant \omega_1^*,\\
            \omega_1|V_1|,\quad \omega \geqslant \omega_1^*,
         \end{array}\right.
\eqno{(6.10a)}
$$
а график сдвига фазы (см. (рис. 4)) силы трения -- по формуле
$$
\varphi=\left\{\begin{array}{c}
           \arg[i(\eta_0-V_1)]-\pi,\quad 0 \leqslant \omega_1 \leqslant \omega_1^*,\\
            \arg[iV_1],\quad \omega \geqslant \omega_1^*. \end{array}\right.
\eqno{(6.10b)}
$$

Из рис. 4 видно, что $\varphi(0)=-\dfrac{\pi}{4}$, что согласуется с
формулой (24,6) из \cite{LandauG}.

Рассмотрим вопрос о диссипации энергии колеблющейся пластины. Рассмотрим мощность диссипации энергии, т.е. величину диссипации энергии в единицу времени, приходящуюся на единицу площади колеблющейся пластины. Согласно
\cite{LandauE} усредненная по времени мощность диссипации энергии вычисляется по формуле
$$
W=\dfrac{1}{2}\Re\Big(u_0F_0^*\Big)=\dfrac{U_0}{2\sqrt{\beta}}\Re F_0^*.
\eqno{(6.11)}
$$

В формуле (6.11) звездочка ($*$) означает комплексное сопряжение.

Рассмотрим случай нулевого индекса задачи, т.е. $\omega_1>\omega_1^*$.
Тогда величина $F_0$ определяется выражением
$$
F_0=2pU_0\omega_1(iV_1).
\eqno{(6.12)}
$$

\begin{figure}[t]
\begin{center}
\includegraphics[width=15.0cm, height=9cm]{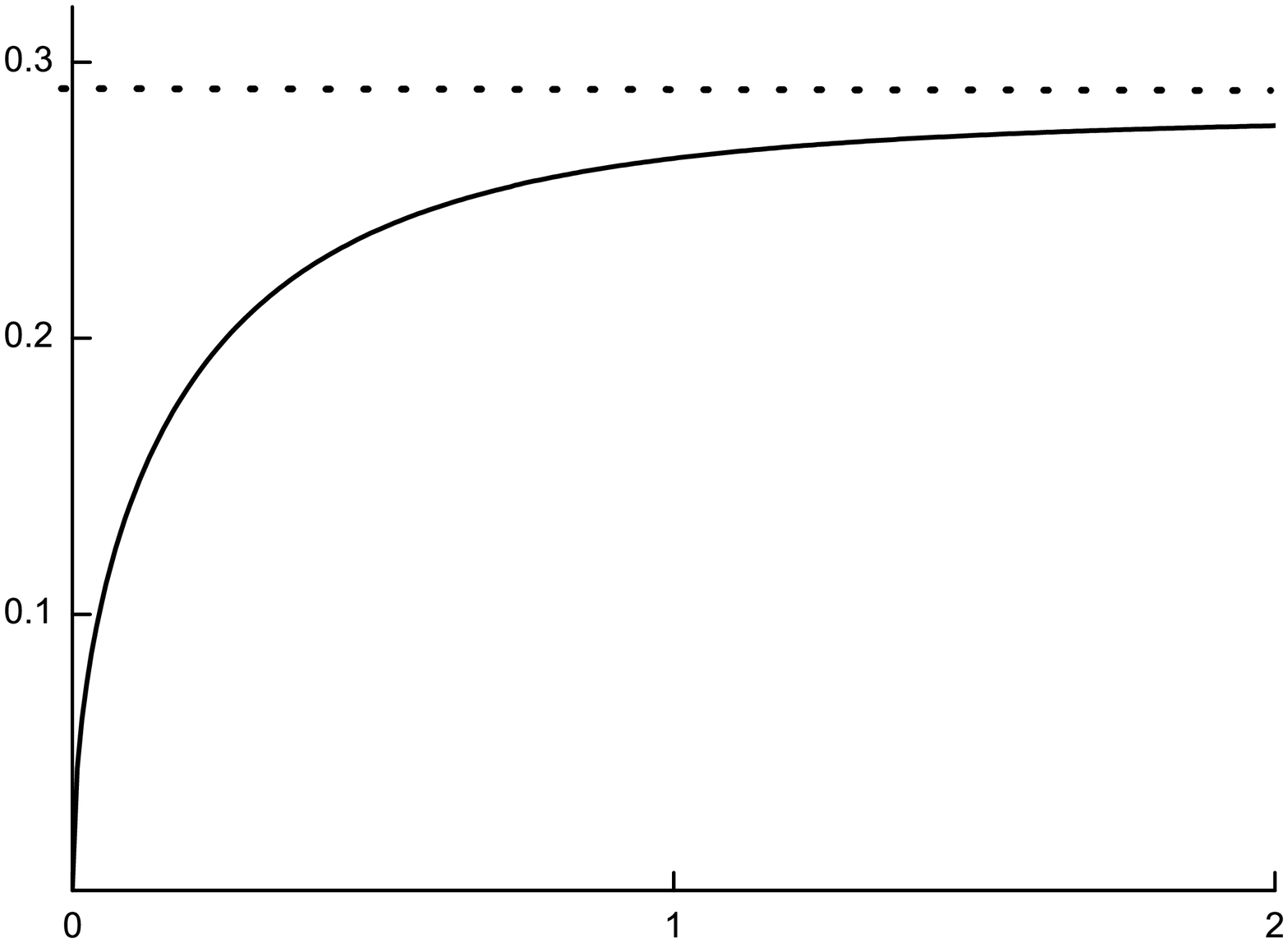}
\end{center}
\begin{center}
{{Рис. 5. Зависимость величины мощности диссипации энергии от частоты колебаний ограничивающей газ плоскости. Эта зависимость при
$0\leqslant \omega_1 \leqslant \omega_1^*$ построена по формуле (6.13), а при $\omega \geqslant \omega_1^*$ -- по формуле (6.14),
$W_0=U_0^2{p}/{\sqrt{\beta}}$.}}
\end{center}
\end{figure}

Из формул (6.11) и (6.12) вытекает, что мощность диссипации энергии на единицу площади пластины равна:
$$
W=U_0^2\dfrac{p}{\sqrt{\beta}}\omega_1\Re(iV_1)^*.
\eqno{(6.13)}
$$

Теперь рассмотрим случай, когда индекс задачи равен единице, т.е.
$\omega_1\in [0,\omega_1)$. В этом случае согласно предыдущему имеем:
$$
W=U_0^2\dfrac{p}{\sqrt{\beta}}\omega_1\Re\Big[i(V_1-\eta_0)\Big]^*.
\eqno{(6.14)}
$$

\begin{center}
\bf 12. Заключение
\end{center}
В настоящей работе сформулирована и решена аналитически вторая задача Стокса --- задача о поведении
разреженного газа, занимающего полупространство над стенкой, совершающей
гармонические колебания. Рассматриваются диффузные граничные условия.
Используется линеаризованное кинетическое уравнение, полученное в результате
линеаризации модельного кинетического уравнения Больцмана в
релаксационном приближении.

На основе аналитического решения найдена скорость разреженного газа в полупространстве и непосредственно у стенки. Отыскивается также сила трения, действующая со стороны газа на пластину, и мощность диссипации энергии пластины.


\makeatother {\renewcommand{\baselinestretch}{1.17}


\begin{thebibliography}{99}\normalsize


\bibitem{Stokes}{\it Stokes G.G.} On the effect of internal friction of fluids on the motion of pendulums. Trans. Cambr. Phil. IX, 8 A851), Math, and Phys. Papers III, 1--141, Cambridge, 1901.

\bibitem{Yakhot}{\it Yakhot V., Colosqui C.} Viscoelastic--Elastic Transition in the "Stokes' Second Problem" in a High Frequency Limit. // arXiv:nlin.CD/0609061.

\bibitem{3}{\it Абрашкин А.А., Якубович Е.И.} Вихревая динамика в лагранжевом описании.-- М.: ФИЗМАТЛИТ; 2006 г.; 175 стр.

\bibitem{4}{\it Шлихтинг Г.} Теория пограничного слоя. М.: Наука, 1974, 712с.

\bibitem{5}{\it Asghar S., Nadeem S., Hanif K., Hayat T.}
 Analytic solution of Stokes second problem for second grade fluid, Math. Probl. Eng. V. 2006, Article ID 72468, 8 p.

\bibitem{6}{\it Ai L., Vafai K.} An Investigation of Stokes' Second Problem for Non-Newtonian Fluids //Numerical Heat Transfer, Part A: Applications, V. 47, 2005, P. 955 - 980.

\bibitem{7}{\it Khan M., Anjum Asia, Fetecau C.} On exact solutions of Stokes second problem for a Burgers' fluid, I. The case $\gamma <
    \lambda^2/4$. // J. Appl. Math. and Phys. (ZAMP). Published online: 26 August 2009.

\bibitem{8} {\it Graebel W.P.} Engineering Fluid Mechanics. New York, Taylor $\&$ Francis, 2001, 676 p.

\bibitem{SS-2002}{\it Siewert C.E., Sharipov F.} Model equations in rarefied gas dynamics: viscous--slip and thermal--slip coefficients // Phys. Fluids. 2002. V. 14, No. 12, 4123-4129.

\bibitem{SK-2007} {\it Sharipov F. and Kalempa D.} Gas flow around a longitudinally oscillating plate at arbitrary ratio of collision frequency to oscillation frequency// Rarefied Gas Dynamics: 25-th International Symposium, edited by M.S.Ivanov and A.K.Rebrov. Novosibirsk, 2007. P. 1140-1145.

\bibitem{10}{\it Karabacak D.M., Yakhot V., and Ekinci K.L.} High--Frequency Nanofluidics: An Experimental Study using Nanomechanical Resonators, Phys. Rev. Lett. 98, 254505, 2007.

\bibitem{11} {\it Cleland A.N., Roukes M.L.} Ananometre--scale mechanical electrometer // Nature, vol. 392, 1998, p. 160-162.

\bibitem{12}{\it Steinhell E., Scherber W., Seide M., Rieger H.}
Investigation on the interaction of gases and well defined solid surfaces with respect to possibilities for reduction of aerodynamic friction and aerothermal heating // Rarefied gas dynamics. Ed. J.L. Potter. N.Y.: Acad. press, 1977. P. 589-602.


\bibitem{13} {\it Дудко В.В., Юшканов А.А., Яламов Ю.И.} Влияние свойств
поверхности на характеристики сдвиговых волн// ЖТФ. 2005. Т. 75, вып.4, 134-135.


\bibitem{14} {\it Дудко В.В., Юшканов А.А., Яламов Ю.И.} Генерация колеблющейся поверхностью сдвиговых волн в газе// ТВТ. 2009. Т. 47.
No. 2, 262-268.

\bibitem{15} {\it Дудко В.В. } Скольжение разреженного газа вдоль
неподвижных и колеблющихся поверхностей, дисс., Москва, 2010. 108 стр.

\bibitem{ALY-1} {\it Akimova V.A., Latyshev A.V., Yushkanov A.A.}
Analytical solution of the second Stokes problem on behaviour of gas over
oscillation surface. Part I: eigenvalues and
eigensolutions//ArXiv: 1111.3429v1 [math-ph] 15 Nov 2011, 27 pp.

\bibitem{ALY-2} {\it Akimova V.A., Latyshev A.V., Yushkanov A.A.}
Analytical solution of the second Stokes problem on behaviour of gas over
oscillation surface. Part II: mathematical apparatus of solving of problem//ArXiv: 1111.5182v1 [math-ph] 22 Nov 2011, 26 pp.

\bibitem{16} {\it Черчиньяни К.} Теория и приложения уравнения Больцмана,
К. Черчиньяни - М.: Мир, 1978.

\bibitem{17} {\it Жаринов В.В., Владимиров В.С.}
Уравнения математической физики, М.: Физмалит, 1999.

\bibitem{18} {\it Латышев А.В., Юшканов А.А.}
Аналитические методы в кинетической теории, Монография. Изд-во МГОУ, М., 2008, 280 с.

\bibitem{Gakhov} {\it Гахов Ф.Д.} Краевые задачи. М.: Наука, 1987, 677 с.

\bibitem{19} {\it Latyshev A.V., Yushkanov A.A.}
Skin effect with arbitrary specularity in Maxwellian Plasma// J. of Math.
Phys. 2010. V. 51, P. 113505-1-113505-10, pp. 10.

\bibitem{20} {\it Latyshev A.V., Yushkanov A.A.}
 Temperature jump in degenerate quantum gases with the Bogoliubov
 excitation energy and in the presence of the Bose - Einstein condensate
 // Theor. and Mathem. Physics, 165(1): 1359 - 1371 (2010).

\bibitem{LandauE} {\it Ландау Л.Д., Лифшиц Е.М.}
Электродинамика сплошных сред. Теоретическая физика. Т. VIII. М. Физматлит (2003), 656 с.

\bibitem{LandauG} {\it Ландау Л.Д., Лифшиц Е.М.}
Гидродинамика. Теоретическая физика. Т. VI. М. Физматлит (1987), 735 с.

\end{thebibliography}
\end{document}